\documentclass[aps,prb,twocolumn,superscriptaddress,nofootinbib,floatfix,amsmath,amssymb]{revtex4-2}

\usepackage[utf8]{inputenc}
\usepackage{amsmath}
\usepackage{slashed}
\usepackage{graphicx}
\usepackage[caption=false]{subfig}
\usepackage{xcolor}
\usepackage[pdftex]{hyperref}
\begin{document}
\title{Derivation of the deformed Heisenberg algebra from discrete spacetime}
\author{Naveed Ahmad Shah}
\affiliation{Department of Physics, Aligarh Muslim University, Aligarh- 202002, India}
\author{S. S. Z. Ashraf}
\affiliation{Department of Physics, Aligarh Muslim University, Aligarh- 202002, India}
\author{Aasiya Shaikh}
 \affiliation{Design and Manufacturing Technology Division, Raja Ramanna Centre for Advanced Technology (RRCAT), Indore-452013, Madhya Pradesh, India}		
\author{Yas Yamin}
\affiliation{Irving K. Barber School of Arts and Sciences,
University of British Columbia - Okanagan, Kelowna,
British Columbia V1V 1V7, Canada}	
\author{P.K. Sahoo }
\affiliation{Department of Mathematics, Birla Institute of Technology and
Science-Pilani,\\ Hyderabad Campus, Hyderabad-500078, India.}
\author{Aaqid Bhat}
 \affiliation{Department of Mathematics, Birla Institute of Technology and
Science-Pilani,\\ Hyderabad Campus, Hyderabad-500078, India.}	
\author{Suhail Ahmad Lone}
\affiliation{Department of Physics and Astrophysics, University of Delhi, Delhi 110007, India}
\author{Mir Faizal}
\affiliation{Irving K. Barber School of Arts and Sciences,
University of British Columbia - Okanagan, Kelowna,
British Columbia V1V 1V7, Canada}
\affiliation{Canadian Quantum Research Center, 204-3002, 32 Ave Vernon, British Columbia, V1T 2L7 Canada}
\affiliation{Faculty of Sciences, Hasselt University, Agoralaan Gebouw D,  Diepenbeek, 3590 Belgium}
\affiliation{Department of Mathematical Sciences, Durham University,
Upper Mountjoy, Stockton Road, Durham DH1 3LE, UK}
\author{M. A. H. Ahsan}
\affiliation{Department of Physics, Jamia Millia Islamia, New Delhi - 110025, India}	
%\date{\today}

\begin{abstract}
Although the deformation of the Heisenberg algebra by a minimal length has become a central tool in quantum gravity phenomenology, it has never been rigorously obtained and is often derived using heuristic reasoning. In this study, we move beyond the heuristic derivation of the deformed Heisenberg algebra and explicitly derive it using a model of discrete spacetime, which is motivated by quantum gravity. Initially, we investigate the effects of the leading order Planckian lattice corrections and demonstrate that they precisely match those suggested by the heuristic arguments commonly used in quantum gravity 
phenomenology. Furthermore, we rigorously obtain deformations from the higher-order Planckian lattice corrections. In contrast to the leading-order corrections, these higher-order corrections are model dependent. We select a specific model that breaks the rotational symmetry, as the importance of such rotational symmetry breaking lies in the relationship between CMB anisotropies and quantum gravitational effects. Based on the mathematical similarity of the Planckian lattice used here with the graphene lattice, we propose that graphene can serve as an analogue system for the study of quantum gravity. Finally, we examine the deformation of the covariant form of the Heisenberg algebra using a four-dimensional Euclidean lattice.
\end{abstract}

\maketitle
\noindent
\textbf{\emph{Introduction.}} There are strong indications that the Heisenberg algebra is deformed due to the quantum gravitational effects  \cite{ALI2009497,DAS2010407,
ALI201184,DAS2008221301}. Even though this deformation has become  the main tool used to study the   quantum gravitational modifications in various systems \cite{DAS2008221301,PIKOVSKI2012393,KHODADI20181659}, the deformation  itself is based on many heuristic  arguments and has never been rigorously derived \cite{KEMPF19951108,TAWFIK2015126001}.  
In constructing the heuristic argument generalized uncertainty principle is related to the existence of a minimal length, which occurs in any approach to quantum gravity \cite{DAS2008221301,PIKOVSKI2012393,KHODADI20181659}. 
The existence of such a minimal length is argued using black hole physics, making it impossible to probe any length below the Planck scale \cite{MAGGIORE199365,PARK2008698}.  To reconcile quantum gravitational minimal length with foundations of quantum mechanics, the usual uncertainty principle is modified to a generalized uncertainty principle  \cite{KEMPF19951108,TAWFIK2015126001}.   A modification of the uncertainty principle also results in the modification of the Heisenberg algebra \cite{ALI2009497,DAS2010407,
ALI201184,DAS2008221301}. It is this modified Heisenberg algebra, which has been used to determine the phenomenological consequences  of quantum gravitational effects on varied physical systems \cite{DAS2008221301,PIKOVSKI2012393,KHODADI20181659}. However, the derivation of the deformed Heisenberg algebra is only heuristic, with no explicit derivation from any model of quantum gravity theories. Besides, a fundamental problem with the heuristic derivation of the algebra is that it can even occur in classical theories \cite{BOSSO2018126010}. Thus, it is not clear if it can have its  origins in the uncertainty principle. We thus shed light on this issue deriving the deformation of Heisenberg algebra starting from discrete spacetime. This will be explicitly done in this paper using a specific model of quantum gravity, called quantum graphity, where the underlying spacetime is discrete and has a particular structure. Various models of quantum gravity, such as causal set theory \cite{BOMBELLI1987521},   loop quantum gravity \cite{ACHOUR2014064064},  multifractal spacetime \cite{CALCAGNI2011061501}, and spin foam  \cite{Asante2020} suggest that spacetime is discrete. Discrete models like the Ising model have been used in quantum gravity, where the spacetime emerges from a Planckian Ising lattice \cite{Feller2016,  Holm1995}.  In quantum graphity, spacetime points are represented by nodes of a hexagonal graph, connected by links that can be on or off. This model thus maps onto a model of discrete spacetime, with honeycomb lattice structure \cite{Quach2012,Konopka2008}. The advantage of taking a honeycomb packing for discrete spacetime, rather than an 
Ising model  \cite{Feller2016,Holm1995} is that in this model it is  naturally possible to account for the breaking of isotropy  of space, which could potentially explain the CMB anisotropies \cite{Land2005,Braglia2021,Acharya2022,Greco2022}. This is due to the close relation between a Planckian honeycomb lattice and  dynamical triangulation \cite{Ito2022,Dittrich2012,Clemente2018,Ambjorn2016,Rindlisbacher2015}. The honeycomb lattice has also been used in  a model of quantum gravity where a  quantum Ricci curvature is obtained from   discrete  spacetime   \cite{Klitgaard2018}.

As quantum graphity has been studied using honeycomb structure \cite{Quach2012,Konopka2008}, at large distance the Planck scale can be approximated using continuum field theories  \cite{Wilkinson2014,Wilkinson2015}. 
Mean-field theory has been used to study the properties of spacetime described by quantum graphity at low temperature \cite{Caravelli2011}. It has also been observed in quantum graphity, that Lorentz symmetry emerges as a low energy symmetry of the theory \cite{Caravelli2012}. The 
honeycomb structure  becomes important in models of dynamical triangulation in quantum gravity \cite{Ito2022,Dittrich2012,Clemente2018,Ambjorn2016,Rindlisbacher2015}.  In fact, we will demonstrate that an advantage of  the underlying trigonal symmetry of the  honeycomb lattice is that it can break rotation symmetry, and this  could potentially explain the CMB anisotropies \cite{Land2005,Braglia2021,Acharya2022,Greco2022}. Thus, there are strong motivations from various approaches to quantum gravity to assume that spacetime can be represented by a honeycomb discrete lattice, rather than an Ising model. Here, we will use this model of discrete spacetime to obtain the deformation of Heisenberg algebra. Here, we note that quantum electrodynamics has also been studied on discrete spacetime lattice  \cite{Magnifico2021}.
  It has been argued that the underlying trigonal symmetry of this honeycomb lattice can modify the low energy Heisenberg algebra of graphene \cite{Iorio2022}.
We will also comment on graphene as an analogue for such models of quantum gravity. This is possible due to the fact that the quasiparticles in graphene simulate relativistic Dirac fermions \cite{Katsnelson2007}, and the underlying lattice is discrete. Various phenomena predicted from quantum electrodynamics have been tested  using graphene \cite{Zhang2022,Berdyugin2022}.  Here, we take the correspondence further and explore this possibility of using Dirac materials like graphene as a quantum gravity analogue \cite{Iorio2022} vis-a-vis the symmetry of the discrete spatial lattice.  This correspondence arises in such materials because of the discrete lattice structure. Since it is not possible to probe graphene below the lattice length scale, this lattice length scale acts as an analogous minimal measurable length scale. As a result in an analogous manner to the minimal length scale, it deforms the Heisenberg algebra used to define the Dirac structure in graphene \cite{Shah2022,Iorio2018a,Iorioplb}.

Finally, motivated by such a deformation, we propose a very general deformation of covariant Heisenberg algebra. To properly define and deform the covariant Heisenberg algebra, we have to use a formalism of quantum mechanics, where time is viewed as an quantum observable\cite{Busch1994,Olkhovsky2009,Olkhovsky2007,Brunetti2010,Olkhovsky2008}. Here, we would like to point out that by slightly modifying the foundations of quantum mechanics, it has been possible to use time as a quantum observable. This has then been used to define a minimal time, and a deformation of the commutator involving time  \cite{Faizal2016,Faizal2017}. Here, we will generalize these results by considering a more general deformation of the covariant algebra from quantum gravitational effects.  
The specific forms of deformation, which in turn would depend on the details of the quantum gravitational model, would produce different forms of these corrections. However, the important observation is that the leading order corrections have a universal form, as the previously heuristic augments have suggested \cite{ALI2009497,DAS2010407,
ALI201184,DAS2008221301}. \\
\noindent 
\textbf{ \emph{Deformed Heisenberg Algebra.}} As the matter fields in nature are Fermionic, for  quantum graphity with honeycomb lattice, we assume that the low energy effective field theory obtained from this lattice is a Fermionic field theory. This is consistent with using graphene as an analogue for this model of quantum gravity \cite{Iorio2022,SalgadoRebolledo2023}. 
Now for such Fermionic effective field theory, we can write the dispersion relation using the analogy with graphene.

The derivation of dispersion relation does not depend on the details of  any system, other than an underlying honeycomb lattice, and hence it will hold for the   honeycomb   structure   in  quantum graphity \cite{Quach2012,Konopka2008}. Thus, for the honeycomb Planckian lattice, where we take into consideration both the next and the next-to-nearest lattice site interaction, we can directly write the  energy dispersion relation as  $E  =     (\lambda p - \alpha p^2) $ \cite{Shah2022,Goerbig2011}.
Here, in this Planckian honeycomb lattice, we  observe that in natural units the Planck energy, $ E_P = L_P^{-1}$, acts as an analogue for the lattice constant in Dirac materials like  graphene. Furthermore, the Fermi velocity  is replaced by velocity of light, which is set equal to one in natural units, $c =1$. It is due to this similarity of the modified dispersion relation \cite{Liberati2014,AmelinoCamelia2009},   that graphene has been proposed as a quantum gravity analogue \cite{Iorio2022, SalgadoRebolledo2023}. This similarity becomes explicit in models of quantum gravity, with a Planck scale honeycomb  discrete lattice.  Here, we have defined  a deformation parameter,  $\alpha$, which can be related to the leading order, $T_P$, and the next to leading  order, $T'_P$,    correction  terms for the Planckian lattice. We also define $\lambda = \pm $, where its value depends on the occupancy of the lattice point. Thus, for any model, with honeycomb lattice, we can write $ \alpha = ({3}/{2} E_P) {|T_P^{'}|}/{T_P} $. 

We start with such an underlying discrete structure,  and then obtain the deformation of Heisenberg algebra using it. 
Using the Dirac prescription $p \rightarrow \boldsymbol{\sigma} \cdot   {\boldsymbol{p}}$ in the expression for the energy, where $  {\boldsymbol{p}} = -i \partial $ is now the momentum operator and $\boldsymbol{\sigma}$ are  Pauli matrices. The effective Hamiltonian  can be expressed as ${H} =  \left[ \boldsymbol{\sigma} \cdot   {\boldsymbol{p}} - \alpha (\boldsymbol{\sigma} \cdot   {\boldsymbol{p}})  (\boldsymbol{\sigma} \cdot   {\boldsymbol{p}})\right]. $
From the definition of the standard low-energy momentum operator, we define a high-energy generalized momentum operator $\boldsymbol{  {P}} =   {\boldsymbol{p}} (1 - \alpha   {p})$,  such that $\boldsymbol{\sigma} \cdot \boldsymbol{  {P}} = \boldsymbol{\sigma} \cdot   {\boldsymbol{p}}(1 - \alpha \boldsymbol{\sigma} \cdot   {\boldsymbol{p}})$. The Hamiltonian, in terms of the generalized momentum operator can be written as $ {H} =   \boldsymbol{\sigma} \cdot \boldsymbol{  {P}} $.
Even though the generalized momentum variables ${P}_i$ are functions of ${p}_i$, we can choose $  {X}_i =   {x}_i$, where  $  {X}_{i}$ is the generalized coordinate conjugate  to  
${P}_i$.  As   the low-energy coordinates  and momentum  operators  satisfy the standard commutation relations $\left[  {x}_i ,   {x}_j \right] = \left[  {p}_i ,   {p}_j \right] = 0$, $~\left[  {x}_i,   {p}_j\right] = i     \delta_{ij}$, the 
generalized commutation relations can be written  as  
$ 
[  {X}_i,   {P}_j] %&=&[x_i, p_j(1 - \alpha p)] \nonumber \\
= i     \left[\delta_{ij} - \alpha \left(  {P} \delta_{ij} + \frac{  {P}_i \   {P}_j}{  {P}}\right) \right].
$ 
  The interesting   observation  here is that this deformation, which is obtained rigorously from discrete spacetime, is exactly the same as has been heuristically argued for, using the generalized uncertainty principle \cite{ALI2009497,DAS2010407}.  
Due to the mathematical similarity of the underlying structure of discreteness between this model of quantum gravity and graphene, it is expected that a similar structure should also occur in graphene at leading order. In fact,  a generalized commutation has also been obtained  for graphene by considering next-to-nearest atomic interactions  \cite{Iorio2018a}, and even though it operates at a different scale, it has the same form as the deformation obtained here. This motivates the potential use of graphene as an analogue for such a model of quantum gravity \cite{Iorio2022,SalgadoRebolledo2023}.  Here, we have been able to directly make  the connection between this algebra and  discrete spacetime using the analogy of graphene with honeycomb lattice in quantum graphity. In fact, such an algebra will hold for any effective lattice field theory. 

Since the leading contribution of the high-energy  momentum  can be obtained by neglecting the effect of lattice hopping terms other than that of the nearest lattice sites, we can write it as a sum of the  low-energy  momentum, and a correction term to it. We thus represent the corrections to the low energy momentum as $\Delta p$, expressing the high energy momentum $\boldsymbol{P}$ as $\boldsymbol{P}= \boldsymbol{p}+ \Delta \boldsymbol{p}$. We can now obtain the explicit expression for the correction term $\Delta p$ in terms of $p$ by observing that $
 \Delta \boldsymbol{p} = (-\alpha \ p \ p_x, -\alpha \ p \ p_y).
$
Now because of the symmetric expansion (isotropic transformation on $\boldsymbol{P}$) the rotational symmetry is preserved. The above transformations allow consistent identification of the high-energy generalized momentum as $P_i = p_i \left( 1 - a \ p\right), $   
corresponding to the generalized canonical coordinates $X_{i}$ \cite{ALI2009497,DAS2010407}.
The full $SO(2, 1)$ Lorentz symmetry is broken by virtue of the presence of $ {P_i P_j}/{P}$ term in the commutation relations. Thus the modified Dirac equation corresponding to this modified Dirac Hamiltonian is given by
$
   \left[ - i     \ \sigma_i \cdot \partial_i -  \alpha      \ \partial^{2}_{i} \right] \psi(\boldsymbol{r}) = E \psi(\boldsymbol{r}), \label{mdeq} 
 $
where $\psi(\boldsymbol{r})$ is a two component spinor. It is important to note that identical modification to the Dirac  equation has been obtained from phenomenological considerations in  quantum gravity \cite{Das2010,Pedram2011}.  Here, it is obtained due to the leading order Planckian corrections to the Dirac structure, modeling the discrete spacetime as a continuum at the leading order. 
This deformation indicates that this modification occurs in any system 
with a discrete structure, if we consider the corrections beyond the continuum approximation. The Planckian  structure of quantum graphity is approximated by a continuum Dirac structure at the leading order. Going beyond the continuum approximation, we obtained a modification to the Dirac equation.
Now, having studied  the  next-to-nearest hopping and leading order contributions in quantum graphity, we can investigate the possible modification due to quantum graphity, beyond leading order corrections.\\
\noindent
\textbf{\emph{Rotational symmetry breaking.}}
The advantage of using a rigorous derivation of the deformation is that we can now precisely see how different models of quantum gravity can produce different deformations of the Heisenberg algebra beyond the leading order. The deformation we discussed in the previous section is universal, in the sense that it can be obtained from any model of quantum gravity using the generalized uncertainty principle (GUP), and hence cannot discriminate between different models of quantum gravity \cite{DAS2008221301,
PIKOVSKI2012393,
KHODADI20181659}. However, going beyond that is not possible using the heuristic reasoning based on GUP. To this end, we rigorously obtain the deformation, which explicitly breaks the rotational symmetry of the effective field theory and hence the isotropy of space. 

%This is important as  the deformation obtained in the previous section preserved the isotropy of space due to its rotational symmetry.
The importance of this study lies in the fact that 
%However,  it is known 
that there are   CMB anisotropies \cite{Land2005,Braglia2021,Acharya2022,Greco2022}, and it has been suggested that they could be explained using quantum gravitational corrections  \cite{Koyama2003}. 
 In fact, a deformation of Heisenberg algebra, which breaks isotropy of space has been proposed to address such  phenomenological problems \cite{Mann2021}. 
 However, again this algebra has been obtained using phenomenological considerations, without it being  derived  from any  model of quantum gravity.  Here, we will obtain a deformation of the Heisenberg algebra using the Planckian honeycomb lattice of quantum graphity. 
It may be noted that in any honeycomb lattice   there is an underlying trigonal symmetry, as the lattice can be considered to be composed of two interpenetrating triangular lattices. Thus, such a symmetry will also occur in quantum graphity with honeycomb lattice \cite{Quach2012,Konopka2008}. %This has important implications for graphene 
%Furthermore, the   trigonal symmetry in quantum gravity occur in models with   dynamical triangulation \cite{Ito:2022ycc, Dittrich:2011vz, Clemente:2018czn, Ambjorn:2016cpa,Rindlisbacher:2015ewa}. 
In fact,   this approach  is based on a higher dimensional analogy of such a trigonal lattice.    Thus, in models of discrete spacetime, like dynamical   triangulation \cite{Ito2022,
Dittrich2012,
Clemente2018,
Ambjorn2016,
Rindlisbacher2015}, and quantum graphity with honeycomb lattice \cite{Quach2012,Konopka2008}, 
we expect deformation of low energy effective field theory. By analogy, we also expect that this 
deformation in $(2+1)$ dimensions (if we limit the discussion to spatial lattice in quantum gravity) should resemble the deformation produced in graphene  \cite{Iorio2022}. 

We investigate the effect of the trigonal symmetry of the lattice on the Heisenberg algebra. We do so by going up to third order terms in the expansion in the quasi-momentum $p$, without taking into consideration the energy contributions from the next-to-nearest neighbor lattice point hoppings. The corresponding Hamiltonian up to $\mathcal{O}((E_p^{-1}|p|)^3)$ is given by
$ 
H(p) =    \Big[ \sigma_x \left( p_x - \frac{1}{4 E_p} ( p^2_x - p^2_y)- \frac{1}{8 E^{2}_p} p_x (p^2_x + p^2_y) \right)  
 + \sigma_y \left(p_y + \frac{1}{2 E_P}  p_x p_y - \frac{1}{8 E_p^{2}} p_y(p^2_x + p^2_y) \right) \Big].
$
%[As we will see, the dispersion relation at this order furnishes us the trigonal warping terms  encoding the trigonal symmetry of the underlying quantum graphity  lattice,  breaking the rotational symmetry.]  
%Thus, such terms  could  potentially be used to explain  CMB anisotropies \cite{Land:2005ad,Braglia:2021fxn,Acharya:2022hsr, Greco:2022ufo}. 
We make a general rotational symmetry breaking expansion,  and define now another high-energy momentum $\boldsymbol{Q}_0$ in terms of low energy momentum $\boldsymbol{p}$ as; $\boldsymbol{Q}_0 = \boldsymbol{p} + \Delta \boldsymbol{q}_0$, where we have $\Delta q_0 = (\Delta q_{0,x}, \Delta q_{0,y})$ ;
\begin{eqnarray}
\Delta q_{0,x} &=&  -\frac{1}{4 E_P } ( p^2_x - p^2_y) - \frac{1}{8 E^2_{p}} p_x (p^2_x + p^2_y) \nonumber \\
\Delta q_{0,y} 
&=& 
\frac{1}{2 E_P}  p_x p_y - \frac{1}{8 E^{2}_{p}} p_y(p^2_x + p^2_y)    
\end{eqnarray}
Thus, we define  $ \boldsymbol{Q}_0 = (Q_{0,x}, Q_{0,y})$,  as $  
  Q_{0,x} =  \left(p_x - \frac{1}{4 E_p} ( p^2_x - p^2_y) - \frac{1}{8 E_p^{2}} p^2 p_x\right), $ and $
  \ \ Q_{0,y} = \left(p_y + \frac{1}{2 E_p} p_x p_y - \frac{1}{8 E_p^{2}} p^2 p_y \right).
$ 
%Using the above expansion, From the Using this low energy effective Hamiltonian, we can derive the deformation of the Heisenberg algebra produced by the third order correction terms. To obtain such corrections to the  Heisenberg algebra, we can define  new generalized momentum in terms of the standard low energy momentum. 
With the transformation, along with the identification of the corresponding generalized coordinate with the standard low-energy ones $X^i_{Q_0} = x^i$, we can write the Hamiltonian for the system in terms of this new generalized momentum as
$  {H} =     \boldsymbol{\sigma} \cdot \boldsymbol{Q}_0
$, and the deformation of the Heisenberg algebra realized is given by $
    [x_i, Q_{0,j}] = i F_{ij}(\boldsymbol{Q}_0) , \ \ [x_i, x_j] = 0 = [Q_{0,i}, Q_{0,j}],
$ 
where we have 
\begin{eqnarray}
F_{ij}(\boldsymbol{Q}_0) &= \delta_{ij} \ + \
 \frac{1}{2 E_p} \begin{bmatrix}
    -Q_{0,x} & Q_{0,y} \\
    Q_{0,y} & Q_{0,x}
    \end{bmatrix} \nonumber \\
    &- \frac{1}{2 E_p^{2}} \begin{bmatrix}
    Q^2_{0,x} &  Q_{0,x} Q_{0,y} \\
    Q_{0,x} Q_{0,y} & Q^2_{0,y}
    \end{bmatrix}. \label{commut2}
\end{eqnarray}
Here again we observe that a  similar algebra, which mathematically resembles this algebra, but acts on a different scale,  has been obtained for graphene using tight-binding model \cite{Iorio2022}.
This again points out to the potential use of graphene as an analogue model for quantum gravity.  
The effective  Hamiltonian,  with   next-to-nearest neighbor  hopping up to third order in momentum  (which includes the trigonal warping term),   can be written as 
\begin{eqnarray}
H(p) = & \Big[ \sigma_x \left( p_x - \frac{1}{4 E_p} ( p^2_x - p^2_y)- \frac{1}{8 E_p^{2}} p_x (p^2_x + p^2_y) \right) \nonumber \\
 & + \sigma_y \left(p_y + \frac{1}{2 E_p}  p_x p_y - \frac{1}{8 E_p^{2}} p_y(p^2_x + p^2_y) \right) \nonumber \\ 
   & - \alpha \left( \frac{1}{E_p} (p^2_x + p^2_y)- \frac{1}{2 E_p^{2}} p^3_x + \frac{3}{2 E_p^{2}} p_x p^2_y \right) \Big].
\end{eqnarray}
Now, we define a still more generalized momentum $\boldsymbol{Q}$ in terms of the momentum $\boldsymbol{Q}_0$ as 
$\boldsymbol{Q}_i = \boldsymbol{Q}_{0,i} (1 - \alpha |\boldsymbol{Q}_{0}|)$.
Corresponding to the general momentum $\boldsymbol{Q}_i$,we now have $  {X}_{i}$ as the generalized conjugate coordinate. Since $  {x}_i$ and $  {p}_i$ are the low-energy coordinates and momentum operators, they satisfy the standard commutation relations, as discussed in the last section too. 
With this generalized momentum a more generalized deformation is realized, which is given by 
$ 
[X_{i}, Q_{j}] = i I_{ij}(\boldsymbol{Q}) , \ \ \ \  [X_{i}, X_{j}] = 0 = [Q_{i}, Q_{j}],
$
where the function $I_{ij}$ is given by 
\begin{eqnarray}
&& I_{ij}(\boldsymbol{Q}) = F_{ik}
(\boldsymbol{Q}) \times \label{l} \\ && \left( (1 -  \alpha |\boldsymbol{Q}| -2  \alpha^2 |\boldsymbol{Q}|^2 ) \delta_{kj}  
-  \alpha \frac{Q_k Q_j}{|\boldsymbol{Q}|} (1 + \alpha |\boldsymbol{Q}|) \right). \nonumber
\end{eqnarray}
Here,  $F_{ik}(\boldsymbol{Q})$ is further given by \eqref{commut2}, where instead of $\boldsymbol{Q}_0$ we will now have $\boldsymbol{Q}$.  A similar algebra has been obtained for graphene \cite{Iorio2018}, due to the mathematical similarity of this model with graphene.  Hence, even to study such deformations, we can use graphene as an analogue.  In general, as we would like the deformation to occur at short distances, we expect the leading order contribution of this algebra to coincide with the usual algebra. Thus, the most general form of $I_{ij} (\boldsymbol{Q})$ can be written as 
$
I_{ij}(\boldsymbol{Q}) = \delta_{ij} + \alpha f(\boldsymbol{Q})_{ij}, 
$
where $f(\boldsymbol{Q})_{ij}$ is a suitable tensorial function formed from the momentum $ {Q}_i$. When we neglect the next-to-nearest hopping and only consider the trigonal warping terms, it reduces to Eq. (\ref{commut2}). Furthermore, by considering both the trigonal warping and next-to-nearest hopping to third order, we obtain Eq. (\ref{l}).   
  This algebra breaks the isotropy of space. The breaking of isotropy   has been motivated by quantum gravitational effects \cite{Mann2021}.  In string theory, 
isotropy can be  broken on branes, and these anisotropic branes are  dual to a deformation of super-Yang-Mills theory by a position dependent term \cite{Mateos2011,HosseiniMansoori2019}. The isotropy can also be broken due to a gravitational Higgs mechanism  \cite{Das2018,Das2018a}.  Thus, it is important to generalize this algebra to four dimensions, and use it to motivate a similar deformation produced by quantum gravitational effects. %However, before we generalize this equation to four dimensions, we note that the specific form of $I_{ij} (\boldsymbol{Q})$ depends on the symmetries of Planckian lattice chosen.
We would also like to point out that for the honeycomb Planckian lattice, Dirac materials can be directly used as an analogue system. This makes it possible to use graphene as an analogue to study proposals used to discuss CMB anisotropies \cite{Land2005,Braglia2021,Acharya2022,Greco2022}. 

\noindent
\textbf{\emph{Covariant Deformation.}}
 In the previous sections, we explicitly analyzed the deformation of the Heisenberg algebra from discrete spatial honeycomb  lattice. To construct  a covariant generalization of such algebra, we need to first consistently define the covariant algebra. 
The problem with defining a covariant algebra is that time is not an operator in ordinary quantum mechanics. However, it 
can be argued that time can be taken as an operator with slight modification to the standard quantum mechanics. 
 It is a widely accepted notion that time cannot be expressed as a self-adjoint operator \cite{Pauli2012}. The reason for this is that 
 a  Hamiltonian with a semi-bounded spectrum does not  admit a group 
of shifts which can be generated from  canonically conjugate self-adjoint operators. This limitation can be resolved in the von-Neumann formulation of quantum mechanics. The resolution is based on the observation that quantum mechanics need not be restricted to
self-adjoint operators \cite{Neumann1932}.   In fact, the  momentum operator for a
free particle bounded by a rigid wall 
at (like time)  is also not a self-adjoint operator. It is possible to include such cases if quantum mechanics is based on a maximal Hermitian operator, and this is done in the  von-Neumann formation of quantum mechanics.

Thus, it is possible to consider time as a quantum mechanical observable
\cite{Busch1994,Olkhovsky2009,Olkhovsky2007,Brunetti2010,Olkhovsky2008}. 
It has also been proposed that time can be investigated using its   symmetric non-self-adjoint operators which  satisfy  
$
[t, H] = - i
$ \cite{Holevo2011,Griffiths2004}. 
Here, observables 
are positive operator valued  measures. 
If $H$ is the Hamiltonian of a system, then a unitary representation of the time translation group can be written as  $\tau \to e^{iH\tau}$.  A quantum mechanical time observable of the system
is represented by  the positive operator valued $B$, such that  $t \to B(t) $. Here, we have  $ e^{iH\tau} B(t) e^{-iH\tau} = B (t - \tau)$. So, a symmetric time operator can be defined for a time observable $B$, such that 
\begin{equation}
   t = \int t \ dB(t).
 \end{equation}
The probability measure $t \to p(t)$ in  any experiment  can be expressed as  $
 p(t) = trace [\rho B(t)].
$
 where  $\rho$  in the density matrix of a  state, and 
$t \to B(t)$ is a positive operator valued  measure \cite{Busch1994}. Thus, the time can be represented as an 
observable in quantum mechanics.

We will use this definition of time to define and deform the covariant commutators.  In fact, such a  deformation of the commutator  which corresponded to a minimum measurable time interval   can be expressed as  \cite{Faizal2016,
Faizal2017}. 
Here, the   observable time is defined   with    reference to the evolution of a
non-stationary quantity. The events were characterized 
by specific values of this quantity
\cite{Busch1994}, and then the minimal value of time is defined with respect to specific values of this quantity.  
For example, for   tunneling time for particles,  a minimum measurable time interval could be viewed as a lower bound on  the measurement of tunneling time for particles  \cite{Busch1994,Olkhovsky2009,Olkhovsky2007,Brunetti2010,Olkhovsky2008}.   This deformed covariant algebra \cite{Faizal2016,
Faizal2017}, with a minimal time  has been related to time crystals, which is the temporal analogy of spatial crystals \cite{Wilczek2012,Shapere2012}. 
 
 It is possible to define a spacetime lattice structure, and this can be done by  wick rotating  time in the complex plane   $\tau = -it$ \cite{Weingarten1982}. Now we can assume that any continuum field theory on  four dimensional Euclidean structure is a low energy approximation to a discrete model of spacetime, with wick rotated time.  Motivated by the previous sections, we can also propose that this Euclidean spacetime 
is represented by a four dimensional generalization of the honeycomb lattice.   Thus, we   propose that a full deformation of a covariant Heisenberg algebra can be written as 
\begin{eqnarray}
[X_{\mu}, K_{\nu}] = i I(\boldsymbol{K})_{\mu\nu},  [R_{\mu}, R_{\nu}] = 0 = [K_{\mu}, K_{\nu}],\label{m}   
\end{eqnarray}
where  $ K_\mu$ is the four momentum, and $I(\boldsymbol{K})_{\mu\nu}$  is given by 
$
I(\boldsymbol{K})_{\mu\nu}  = \delta_{\mu\nu} + \alpha f(\boldsymbol{K})_{\mu\nu}.
$
Here,  $f(\boldsymbol{K})_{\mu\nu}$ is a suitable tensorial function formed from the four  momentum $ {K}_\mu$, such that it 
 respects the symmetries of the Planck scale lattice structure produced by quantum gravity. For example, if we consider dynamical triangulation or quantum graphity with honeycomb lattice, and neglect the temporal deformation of $(2+1)$ dimensions gravity, then this tensorial function in Eq. (\ref{m})  resembles the tensorial function of  Eq. (\ref{l}), with the parameter $\alpha$ being of order of Planck length. Now for  quantum graphity  with four dimensional covariant  honeycomb lattice and    dynamical triangulation in four dimensions, we would expect a deformation in Eq. (\ref{m}),  which would break  the Lorentz  symmetry, like the breaking of isotropy of space by the deformation produced by third order correction in Eq. (\ref{l}). In the limit, higher order correction terms, like  the trigonal warping terms, are neglected, and only the next-to-nearest hopping is considered,  the tensorial function $f(\boldsymbol{K})_{\mu\nu}$ can be expressed as 
$
f(\boldsymbol{K})_{\mu\nu}  = \left(  {K} \delta_{\mu\nu} + \frac{  {K}_{\mu} \   {K}_{\nu} }{  {K}}\right).
$
This deformation of the low energy Heisenberg algebra will occur due to most four dimensional lattice structures. Hence, this deformation has been motivated by the general consideration using quantum gravitational phenomenology  \cite{Faizal2016,
Faizal2017}. However, the deformation produced by higher order correction terms, will depend on the symmetry of  the underlying theory. 
It is possible to study  different  deformations of the covariant Heisenberg algebra, which can correspond to  low energy limits of different quantum theories of gravity. Here, it is important to note that any symmetry like Lorentz symmetry will break due to higher order Planckian lattice corrections, and these symmetries will be restored if these higher order Planckian lattice corrections are neglected.  As these corrections become important near Planck energy, at    low energies comparable to Planck energy, when we neglect all higher order corrections, we obtain the original covariant  Heisenberg algebra.  Such breaking of Lorentz symmetry due to quantum gravitational effects have already been studied in quantum gravity  phenomenology \cite{Delhom2022,Gubitosi2010,Jacob2010,Ding2020}. 
Here, we have investigated the reason for such breaking using a four dimensional Euclidean Planckian  lattice.\\
\noindent
\textbf{\emph{Conclusion.}}
The main result of the present letter is that we have for the first time explicitly obtained the deformation of the Heisenberg algebra using a model of quantum gravity. We observed that the leading order deformation is exactly the same as suggested previously using heuristic arguments based on the generalized uncertainty principle. To obtain these results, we have analyzed low energy consequences of quantum graphity with Planck scale honeycomb 
lattice. As the matter fields are fermionic, we assumed that the low energy effective field theory of this lattice is also a fermionic field theory. 
We first analyzed the effects of leading order discrete Planckian lattice corrections on energy. It was observed that the modifications produced by these are similar to the modifications proposed using quantum gravitational phenomenology. They also resembled the modifications to the Dirac equation in graphene from the leading order correction terms.
%and as such graphene can be used as an analogue system to study it \cite{Acquaviva:2022yiq, Salgado-Rebolledo:2021zbxx}. 
Next, we investigated the effects of this discrete Planckian lattice up to the higher order energy correction terms and obtained a general deformation of the  Heisenberg algebra, which can be reduced to specific forms of the algebra in different limits.  To discern covariant deformed algebra, we used a formalism of quantum mechanics, where time could be taken as an observable. The minimal time restricts the measurement of this observable. We also analyzed various deformation of this covariant algebra. 

It is important to note that the conclusions obtained here may also hold in theories like string theory.  It has been demonstrated that geometries where there is a limit on the scale at which the spacetime can be probed have the same information as discrete spacetime \cite{Kempf2009}. Thus, even for string theory, it can be argued based on the results from information theory \cite{Kempf2009},  that discrete spacetime  could capture some important feature of string theory. Furthermore, it has been demonstrated that spacetime  geometries can emerge on boundaries of tensor networks  \cite{Chen2016}. Here, it is interesting to note that holography has been used to relate discrete spacetime described by  tensor networks   to string theory  \cite{Kibe2022,Jahn2021}. The honeycomb lattice has also been studied using holography in string theory \cite{Das2017,Ryu2010}.  It could thus be possible to holographically obtain the deformation of the Heisenberg algebra. In this regard, the leading order  deformation of the Heisenberg algebra has been obtained  using  holography \cite{Faizal2017,Faizal2015}. However, this has not been done for next to the leading order corrections, which will be different for  theories with honeycomb lattice structure \cite{Das2017,Ryu2010}

There have been several proposals put forward for the detection of the deformation of Heisenberg algebra by Planck scale quantum gravitational effects in varied systems such as macroscopic harmonic oscillators \cite{Bawaj2015}, gravitational bar detectors \cite{Marin2013},  optomechanics \cite{Aghababaei2024}. It was also suggested that the modification of Heisenberg algebra can be detected using ultra-precise measurements of Lamb shift and Landau levels  \cite{DAS2008221301}.  However, due to the feeble nature of the quantum gravitational effects, the corrections are too small to evade experimental detection. As an alternative, condensed matter systems have been used to simulate both classical and quantum gravitational scenarios \cite{Braunstein2023}. 
Graphene, as we discussed is one such candidate which can be used to simulate the discrete spacetime effects on quantum electrodynamic phenomena like Klein tunneling, Schwinger Mechanism etc. Moreover, as we highlighted in the present work, the models of quantum gravity like quantum graphity with honeycomb lattice share some similarities to the tight binding model for graphene.  Thus, 
we propose that graphene can be used as an analogue for quantum gravity and can be used to gain insights into the fundamental nature -- e.g. symmetry-- of the spacetime itself. 

The immediate interesting case of study of the non-linear Hamiltonians considered in the present work is the calculation of transmittance for angular Klein tunneling in graphene. Since the Hamiltonian takes into consideration the higher energy terms in the form of both the next-to-nearest neighbor atomic hopping and trigonal warping terms, this model allows us to predict the corrections to the Klein tunneling at angular incidences to a much better accuracy. Further, it is expected that going to slightly higher energies the corrections occurring from the trigonal warping terms on the angular Klein tunneling would show some appreciable effect on the angular dependence of the transmittance in experiment.
 These corrections would thus correspond to the signatures of anisotropy of space in addition to its discrete topology on the phenomena of Klein tunneling for Dirac fermions. Thus, from the point of view of graphene, the analytical model at higher energies can be used to study the transport phenomena in graphene over a wide energy regime and to better accuracy, for example, angular Klein tunneling \cite{Zhang2022} and the mesoscopic Schwinger effect, which has recently been experimentally detected in graphene \cite{Schmitt2023}. This is true for many other quantum electrodynamic phenomena that have been proposed to occur in graphene.
\acknowledgments
\noindent
N.A.S. acknowledges the Council of Scientific and Industrial Research (CSIR), Government of India, for funding through the CSIR Research Associateship (09/0112(18208)/2024-EMR-I). The authors thank James Quach for discussions related to quantum graphity. The authors also thank Salman Sajad Wani, Alonso Contreras Astorga, and Fran\c{c}ois Fillion-Gourdeau for useful discussions about graphene.  

\end{document}